\documentclass[journal]{IEEEtran}

\usepackage{amsmath,amsfonts}
\usepackage{algorithmic}
\usepackage{algorithm}
\usepackage{array}
\usepackage[caption=false,font=normalsize,labelfont=sf,textfont=sf]{subfig}
\usepackage{textcomp}
\usepackage{stfloats}
\usepackage{url}
\usepackage{verbatim}
\usepackage{mathtools}
\usepackage{graphicx}
\usepackage{subfig} 
\usepackage{bm}
\usepackage{amssymb}
\usepackage{enumitem}
\usepackage[hidelinks]{hyperref}
\usepackage{graphicx}
\usepackage{multirow} 
\usepackage{makecell}  
\usepackage[justification=centering]{caption}
\usepackage{booktabs}
\usepackage{bm}

\renewcommand\eqref[1]{(\ref{#1})}
\usepackage{cite}
\usepackage{xcolor} 

\ifCLASSINFOpdf

\else

\fi

\begin{document}

\title{Dynamic Antenna Placement for Mobile Users in Urban Micro Pinching-Antenna Systems}

\author{Qiushi Zhao, Zihan Feng, Ximing Xie,~\IEEEmembership{Member,~IEEE}, Hao Qin,~\IEEEmembership{Member,~IEEE}, Yuanwei~Liu,~\IEEEmembership{Fellow,~IEEE}, and Xingqi Zhang,~\IEEEmembership{Senior Member,~IEEE}
\thanks{Qiushi Zhao, Zihan Feng, and Hao Qin are with the School of Electronics and Information Engineering, Sichuan University, Chengdu, China (hao.qin@scu.edu.cn).}
\thanks{Ximing Xie is with the Department of Electrical and Computer Engineering, Western University, London, ON N6A 3K7, Canada (xxie269@uwo.ca).}
\thanks{Yuanwei Liu is with the Department of Electrical and Electronic Engineering, The University of Hong Kong, Hong Kong (e-mail: yuanwei@hku.hk).}
\thanks{Xingqi Zhang is with the Department of Electrical and Computer Engineering, University of Alberta, Canada T6G 2H5 (e-mail: xingqi.zhang@ualberta.ca).}
}

\maketitle

\begin{abstract}
The pinching-antenna systems (PASS) enable blockage mitigation in urban micro (UMi) networks through flexible antenna placement. However, the joint optimization of antenna positions and beamforming precoding is inherently nonconvex and becomes significantly more challenging under user mobility. To address this issue, we propose a bilevel optimization framework for dynamic antenna positioning and beamforming precoding design. In the outer level, a soft actor-critic (SAC) agent learns a continuous control policy for real-time antenna positioning, while in the inner level, zero-forcing (ZF) precoding is applied based on the instantaneous effective channel. Numerical results demonstrate that the proposed framework significantly improves spectral efficiency (SE) and enhances robustness against user mobility and random blockages.
\end{abstract}


\IEEEpeerreviewmaketitle

\section{Introduction}
As a promising architecture in the domain of flexible antenna systems, the pinching-antenna system (PASS) offers significant potential for overcoming propagation challenges in sparse line-of-sight (LoS) conditions. The core concept of PASS involves deploying pinching antennas (PAs) on dielectric waveguides, enabling adjustable antenna positions and dynamic reconfiguration of the wireless channels\cite{guo2025gpass}. This geometric flexibility enables the system to actively bypass blockages and establish stable LoS links \cite{ding2025flexible}, which is critical for maintaining reliable connectivity in blockage-rich environments.

In PASS, optimizing antenna positions is indispensable for maximizing system performance \cite{guo2025gpass}. However, prior research in this domain has predominantly focused on static or quasi-static scenarios. Specifically, existing studies have explored maximizing array gain via inter-antenna spacing \cite{ouyang2025array} and discrete position selection \cite{wang2025antenna}, or optimizing deployment for fixed users \cite{zhao2025pinching}. Various algorithms have been proposed, including the non-real-time gradient meta-learning joint optimization (GML-JO) algorithm \cite{zhou2025gradient}, the robust particle swarm optimization (PSO) algorithm \cite{zeng2025robust}, and low-complexity methods for non-orthogonal multiple access (NOMA) systems \cite{xie2025low}. These approaches generally rely on offline optimization designed for static environments. Such static assumptions are fundamentally incompatible with dynamic urban micro (UMi) downlink scenarios, where rapid user mobility causes fast time-varying channel conditions.

Deep reinforcement learning has emerged as a powerful paradigm for handling dynamic optimization complexities in wireless networks \cite{qin2025physics, zhang2021cognitive, wu2025intelligent}. Leveraging this capability, we propose a novel dynamic bilevel optimization framework for the real-time scenario. To decouple the complex optimization variables, we design a bilevel architecture: the outer loop employs a soft actor-critic (SAC) agent to learn the optimal positions of the pinching antennas based on user mobility and channel states, while the inner loop executes a model-based zero-forcing (ZF) algorithm to calculate the optimal transmit beamforming weights. This design leverages the theoretical optimality of ZF for the inner loop while utilizing the SAC algorithm to learn the underlying physical characteristics of the UMi channel environment, and thus achieves precise, real-time optimization. Numerical results demonstrate that the proposed architecture significantly outperforms random and fixed baselines, maintaining robustness in random channel conditions.

\section{System Model and Problem Formulation}

\begin{figure}[htbp]
    \centering
    \includegraphics[width=0.9\linewidth]{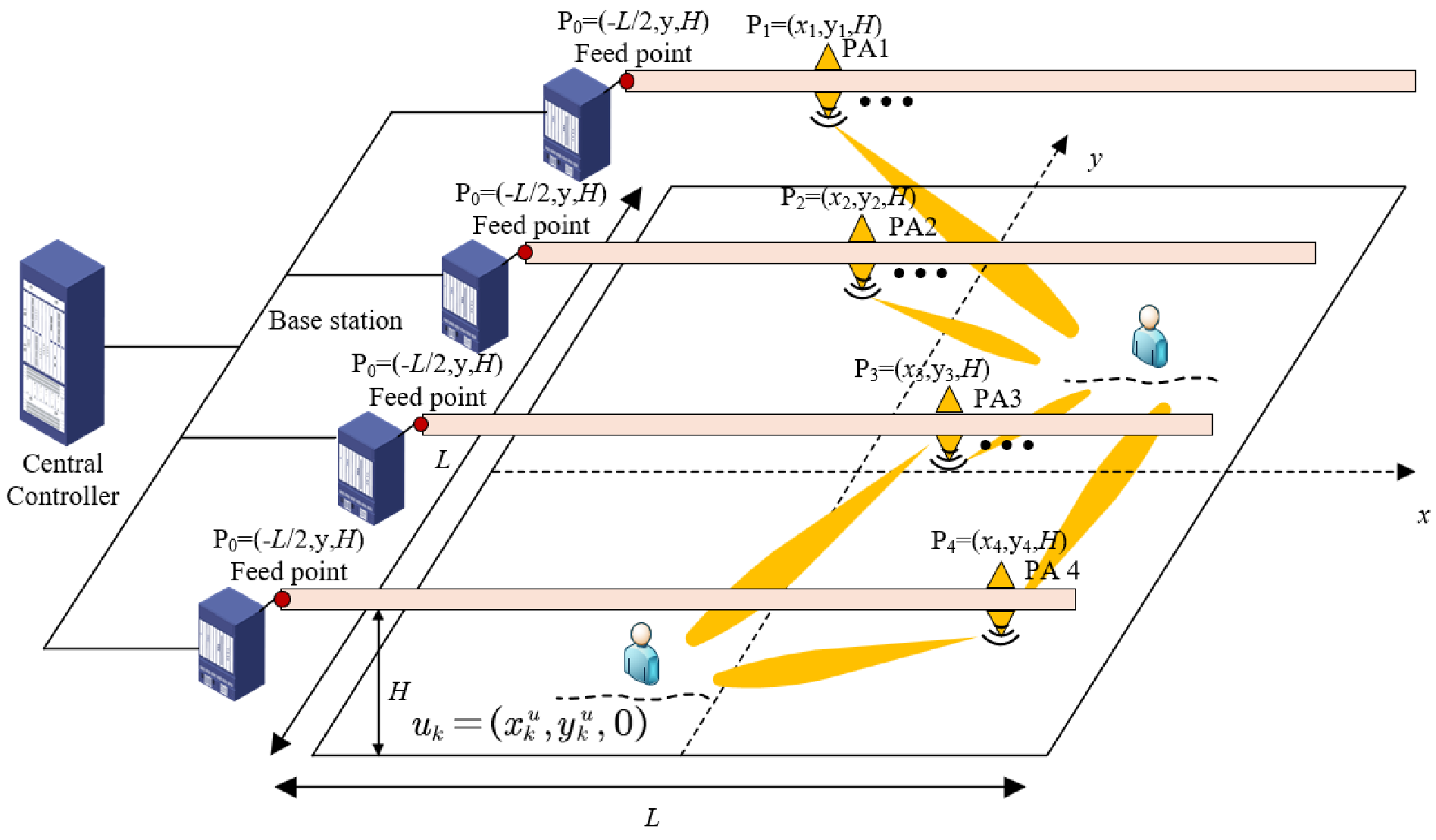} 
    \caption{Illustration of the downlink PASS system with $N$ waveguides and $K$ mobile users in a UMi scenario.}
    \label{fig:system_model}
\end{figure}

\begin{figure*}[htbp]
    \centering
    \includegraphics[width=0.9\linewidth]{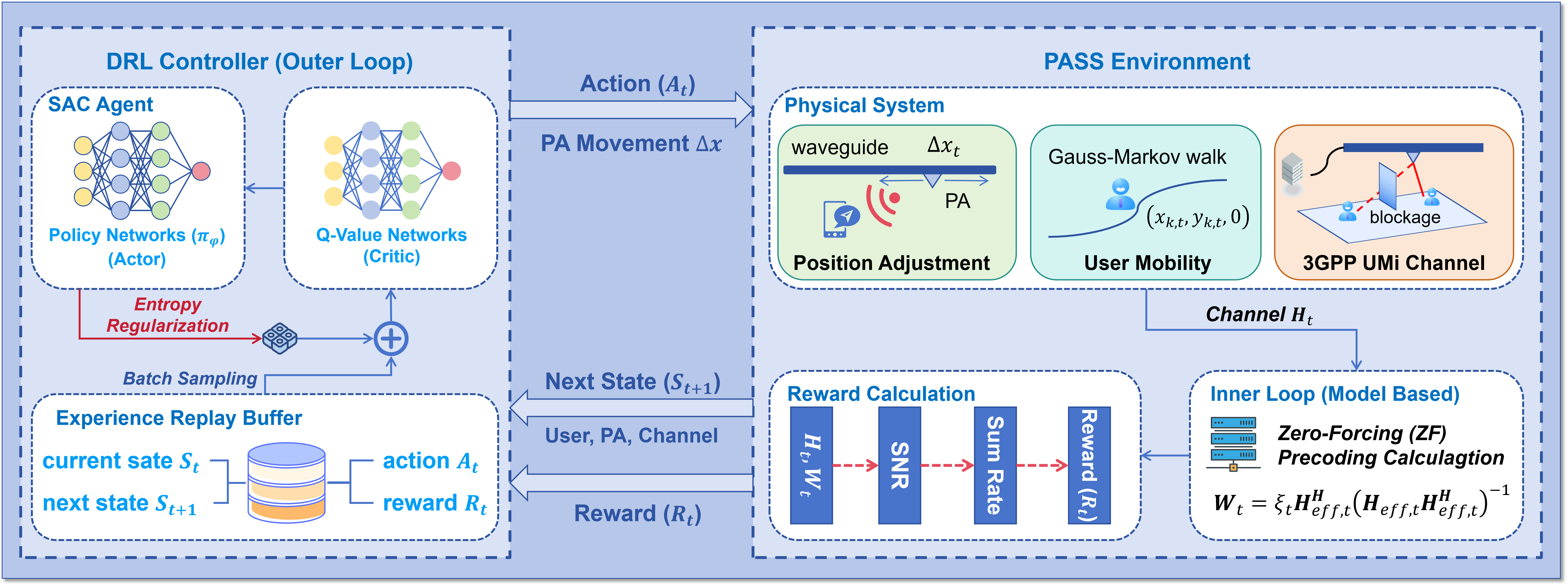} 
    \caption{Framework of proposed bilevel beamforming optimization system in UMi.}
    \label{fig:framework}
\end{figure*}

\subsection{System Scenario and User Mobility}
Consider a downlink communication system deployed in an UMi street canyon scenario, as illustrated in Fig. \ref{fig:system_model}. The system consists of a central controller (CC) and four base stations (BSs). The whole system is equipped with $N$ parallel dielectric waveguides deployed at a fixed height $H$. The waveguides are aligned along the $x$-axis, and their $y$-coordinates are denoted by the set $\mathcal{Y}_{\text{WG}} = \{y_1, \dots, y_N\}$. Each waveguide is equipped with one movable PA. The position of the PA on the $n$-th waveguide at time slot $t$ is denoted as $x_{n,t}$, where $0 \le x_{n,t} \le D$, with $D$ representing the length of the waveguide (and also the side length of the service area).

The system serves $K$ single-antenna users distributed in the area. To capture the temporal correlation of user movement, we adopt a Gauss-Markov mobility model. Let $\mathbf{u}_{k,t} \in \mathbb{R}^2$ and $\mathbf{v}_{k,t} \in \mathbb{R}^2$ denote the horizontal position and velocity vector of user $k$ at time $t$, respectively. The velocity evolves as:
\begin{equation}
    \mathbf{v}_{k,t+1} = \alpha \mathbf{v}_{k,t} + (1-\alpha)\bar{\mathbf{v}} + \boldsymbol{\xi}_{k,t},
\end{equation}
where $\alpha \in [0, 1)$ is the memory level coefficient, $\bar{\mathbf{v}}$ is the asymptotic mean velocity, and $\boldsymbol{\xi}_{k,t}$ is the Gaussian noise vector capturing randomness. The position is updated based on the velocity and the time interval $\Delta T$. This model captures the temporal correlation of human mobility, and is therefore suitable for reinforcement learning optimization. The whole framework of our proposed method is shown in Fig. \ref{fig:framework}.

\subsection{Dynamic Channel Model}
The channel between the $n$-th PA and the $k$-th user is modeled based on the 3GPP TR 38.901 UMi standard\cite{zhu20213gpp}.

\subsubsection{LoS Probability}
The existence of an LoS link depends on the 2D horizontal distance between the $n$-th PA and user $k$. Let $d = d_{k,n,t}^{\text{2D}}$ denotes the 2D distance, the LoS probability $P_{\text{LoS}}$ is given by:
\begin{equation}
    P_{\text{LoS}}(d) = 
    \begin{cases} 
        1, & d \leq 18 \text{ m}, \\[8pt]
        \displaystyle \frac{18}{d} + \exp\left(-\frac{d}{36}\right)\left(1 - \frac{18}{d}\right), & \\[8pt]
          & d > 18 \text{ m}.
    \end{cases}
\end{equation}
To reflect temporal consistency of blockage events, we adopt a first-order Markov model whose stationary distribution matches the 3GPP LoS probability. The actual link state $z_{k,n,t}$ is evolved using a Markov chain based on $P_{\text{LoS}}$.

\subsubsection{Channel Coefficients}
The baseband effective channel coefficient $h_{k,n,t}$ aggregates path loss and small-scale fading:
\begin{equation}
    h_{k,n,t} = \sqrt{\beta_{k,n,t}} \cdot \tilde{h}_{k,n,t},
\end{equation}
where $\beta_{k,n,t} \propto (d_{k,n,t}^{\text{3D}})^{-\eta}$ represents the large-scale path loss with path loss exponent $\eta$ and 3D distance $d_{k,n,t}^{\text{3D}}$. The small-scale fading component $\tilde{h}_{k,n,t}$ is modeled as:
\begin{equation}
    \tilde{h}_{k,n,t} = 
    \begin{cases} 
    e^{-j \frac{2\pi}{\lambda} d_{k,n,t}^{\text{3D}}}, & \text{if } z_{k,n,t} = \text{LoS}, \\
    g_{k,n,t} \sim \mathcal{CN}(0, 1), & \text{if } z_{k,n,t} = \text{NLoS},
    \end{cases}
\end{equation}
where \(\lambda\) denotes the wavelength of the carrier frequency. For tractability, the LoS component is modeled as a dominant deterministic plane wave, and the NLoS case follows a Rayleigh fading distribution.

\subsection{Signal Model and Beamforming}
The system employs a hybrid beamforming architecture. The pinching beamforming matrix $\mathbf{G}_t \in \mathbb{C}^{N \times N}$ represents the phase delay introduced by signal propagation within the waveguides and is expressed as a diagonal matrix:
\begin{equation}
    \mathbf{G}_t = 
    \begin{bmatrix}
    e^{-j \frac{2\pi}{\lambda_g} x_{1,t}} & 0 & \cdots & 0 \\
    0 & e^{-j \frac{2\pi}{\lambda_g} x_{2,t}} & \cdots & 0 \\
    \vdots & \vdots & \ddots & \vdots \\
    0 & 0 & \cdots & e^{-j \frac{2\pi}{\lambda_g} x_{N,t}}
    \end{bmatrix},
\end{equation}
where $\lambda_g$ is the guided wavelength.

The digital transmit beamforming matrix, also known as beamforming precoding at CC, is denoted as $\mathbf{W}_t \in \mathbb{C}^{N \times K}$. The received signal at user $k$ is:
\begin{equation}
    y_{k,t} = \mathbf{h}_{k,t}^{\text{H}} \mathbf{G}_t \mathbf{w}_{k,t} s_{k,t} + \sum_{j \neq k} \mathbf{h}_{k,t}^{\text{H}} \mathbf{G}_t \mathbf{w}_{j,t} s_{j,t} + n_{k,t}.
\end{equation}
where $\mathbf{h}_{k,t} \in \mathbb{C}^{N \times 1}$ is the vector of channel coefficients from all PAs to user $k$, and $n_{k,t} \sim \mathcal{CN}(0, \sigma_n^2)$ is the noise.

Let $\mathbf{H}_{\text{eff}, t} = [\mathbf{G}_t^{\text{H}}\mathbf{h}_{1,t}, \dots, \mathbf{G}_t^{\text{H}}\mathbf{h}_{K,t}]^{\text{H}}$ denote the effective channel matrix seen by the BS. We adopt ZF precoding to mitigate inter-user interference. The precoding matrix is computed analytically as:
\begin{equation}
    \mathbf{W}_t = \zeta_t \mathbf{H}_{\text{eff}, t}^{\text{H}} (\mathbf{H}_{\text{eff}, t} \mathbf{H}_{\text{eff}, t}^{\text{H}})^{-1},
\end{equation}
where $\zeta_t$ is the normalization factor ensuring the total transmit power constraint $\|\mathbf{W}_t\|_F^2 \leq P_{\max}$. We assume $N \ge K$ to ensure that the ZF precoder is well-defined.

The resulting signal-to-noise ratio (SNR) for user $k$ under ZF precoding simplifies to:
\begin{equation}
    \text{SNR}_{k,t} = \frac{|\mathbf{h}_{k,t}^H \mathbf{G}_t \mathbf{w}_{k,t}|^2}{\sigma_n^2}. \label{eq:snr}
\end{equation}

\subsection{Problem Formulation}

We aim to dynamically optimize the PA positions to maximize the long-term spectral efficiency of the system while minimizing the mechanical movement overhead. Let $\boldsymbol{\phi}_t = [x_{1,t}, \dots, x_{N,t}]^T$ denote the vector of PA positions at time slot $t$. The optimization problem is formulated as:

\begin{subequations}
\begin{align}
\text{(P1):} \max_{\{x_{n,t}\}, \forall n, t} \quad & \mathbb{E} \Bigg[ \sum_{t=0}^{\infty} \gamma^t \Bigg( \sum_{k=1}^{K} \log_2(1 + \text{SNR}_{k,t}) \notag \\
& \qquad \qquad - \lambda \sum_{n=1}^{N} |\Delta x_{n,t}| \Bigg) \Bigg] \\
\text{s.t.} \quad & 0 \le x_{n,t} \le D, \quad \forall n, t, \\
& |\Delta x_{n,t}| = |x_{n,t} - x_{n,t-1}| \le \Delta_{\max}, \quad \forall n, t, \\
& \|\mathbf{W}_t\|_F^2 \le P_{\max}, \quad \forall t,
\end{align}
\end{subequations}
where $\gamma \in [0, 1)$ is the discount factor, $\lambda$ is a weighting coefficient balancing the SE and the movement cost, and $\Delta_{\max}$ represents the maximum sliding distance per time slot.

\section{DRL-Based Dynamic Antenna Positioning Design}
\label{sec:drl_algorithm}

We propose an SAC-based framework to solve (P1) by learning a continuous policy $\pi_{\phi}$ that maps observations to PA actions, thereby avoiding redundant optimization at each time slot.

\subsection{MDP Formulation for PASS}
We model the dynamic control of the PASS system as a Markov decision process (MDP) defined by the tuple $(\mathcal{S}, \mathcal{A}, \mathcal{P}, r, \gamma)$. Here, $\mathcal{P}$ represents the state transition dynamics governed by the implicit user mobility and channel evolution, and $\gamma \in [0,1)$ serves as the discount factor for long-term returns. The detailed definitions are as follows:

\subsubsection{State Space $\mathcal{S}$}
To enable the agent to perceive the dynamic environment and make informed decisions, we design a comprehensive 40-dimensional state vector $\mathbf{s}_t$ that aggregates geometric and channel information. Specifically, $\mathbf{s}_t$ consists of:
\begin{itemize}
    \item User Geometry (12 dims): The normalized coordinates $(\tilde{x}_{k,t}, \tilde{y}_{k,t})$ and velocities $(\tilde{v}^x_{k,t}, \tilde{v}^y_{k,t})$ for all $K=3$ users, allowing the agent to predict user mobility trends.
    \item PA Configuration (4 dims): The normalized current positions $\tilde{x}^{\text{PA}}_{n,t}$ of the $N=4$ PAs along the waveguides.
    \item Channel State Information (24 dims): To capture the blockage status without full CSI overhead, we include the normalized channel power gain $\tilde{g}_{k,n,t}$ (in dB) and a binary LoS indicator $l_{k,n,t}$ for each PA-User link.
\end{itemize}
The complete state vector is defined as:
\begin{equation}
\begin{split}
    \mathbf{s}_t = [\underbrace{\dots, \tilde{x}_{k,t}, \tilde{y}_{k,t}, \tilde{v}^x_{k,t}, \tilde{v}^y_{k,t}, \dots}_{\text{Users}}, \underbrace{\tilde{\boldsymbol{\phi}}_t}_{\text{PAs}}, \underbrace{\dots, \tilde{g}_{k,n,t}, l_{k,n,t}, \dots}_{\text{Links}}]^T\\ \in \mathbb{R}^{40}.
\end{split}
\end{equation}

\subsubsection{Action Space $\mathcal{A}$}
The action $\mathbf{a}_t \in [-1, 1]^N$ represents the normalized displacement control for the PAs. The physical displacement $\Delta x_{n,t}$ for the $n$-th PA is mapped as:
\begin{equation}
    \Delta x_{n,t} = a_{n,t} \cdot \Delta_{\max}, \quad n=1,\dots,N,
\end{equation}
where $\Delta_{\max}$ is the maximum mechanical sliding distance per time slot. The PA positions are updated as $x_{n,t+1} = \text{clip}(x_{n,t} + \Delta x_{n,t}, 0, D)$.

\subsubsection{Reward Function $r$}
The reward function is designed to guide the agent toward maximizing the system SE while minimizing mechanical reconfiguration costs. The instantaneous reward $r_t$ is defined as:
\begin{equation}
    r_t = \sum_{k=1}^{K} \log_2(1 + \text{SNR}_{k,t}) - \lambda \sum_{n=1}^{N} |\Delta x_{n,t}|,
\end{equation}
where $\lambda$ balances the trade-off between communication performance and mechanical energy consumption.

\subsection{SAC Algorithm Design}
To prevent local optima in the continuous action space, we employ the SAC algorithm, which maximizes the entropy-regularized objective:
\begin{equation}
    J(\pi) = \sum_{t} \mathbb{E}_{(\mathbf{s}_t, \mathbf{a}_t) \sim \rho_\pi} \left[ r(\mathbf{s}_t, \mathbf{a}_t) + \alpha \mathcal{H}(\pi(\cdot|\mathbf{s}_t)) \right],
\end{equation}
where $\alpha$ is the temperature parameter balancing exploration and exploitation. The framework employs an actor-critic architecture parameterized by
multi-layer perceptrons (MLPs). 

Specifically, we utilize two parallel critic networks (clipped double-Q learning) to mitigate value overestimation. The critics are trained to minimize the mean squared error (MSE) between the predicted Q-values and the soft temporal difference (TD) target, which incorporates both the immediate reward and the entropy-regularized future value. 

Simultaneously, the actor network updates the policy parameters to maximize the expected soft Q-value. To enable gradient updates through the stochastic sampling, the actions are sampled using the reparameterization trick 
\begin{equation}
    \tilde{\mathbf{a}} = \tanh(\mu_\phi + \sigma_\phi \odot \epsilon).
\end{equation}
Implementation details are listed in Table \ref{tab:hyperparameters}.

\begin{table}[htbp]
\centering
\caption{Simulation and Hyperparameter Settings}
\label{tab:hyperparameters}
\begin{tabular}{l c}
\toprule
\textbf{Parameter} & \textbf{Value} \\
\midrule
\textit{System Parameters} & \\
Service Area Size ($D \times D$) & $100 \text{ m} \times 100 \text{ m}$ \\ Signal Frequency & $28GHz$ \\ Number of Waveguides / PAs ($N$) & 4 \\
Number of Users ($K$) & 3 \\
Max PA Sliding Distance ($\Delta_{\max}$) & 3 m \\
Movement Penalty Weight ($\lambda$) & 0.03 \\
\midrule
\textit{DRL Hyperparameters} & \\
Discount Factor ($\gamma$) & 0.99 \\
Replay Buffer Size & 200,000 \\
Batch Size ($B$) & 256 \\
Hidden Layer Units & [256, 256] \\
Learning Rate (Actor/Critic/Alpha) & $3 \times 10^{-4}$ \\
Max Timesteps per Episode & 80 \\
Total Training Episodes & 400 \\
\bottomrule
\end{tabular}
\end{table}

    

\section{Numerical Results}
\label{sec:results}

\begin{figure}[htbp]
    \centering
    \includegraphics[width=0.9\linewidth]{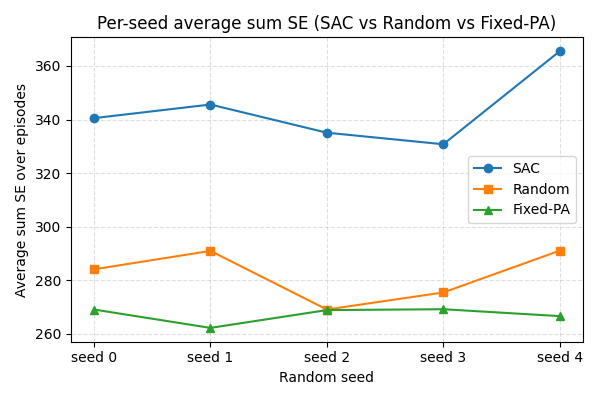}
    \caption{Robustness analysis across 5 different random seeds.}
    \label{fig:seeds}
\end{figure}

\begin{figure}[htbp]
    \centering
    \includegraphics[width=0.9\linewidth]{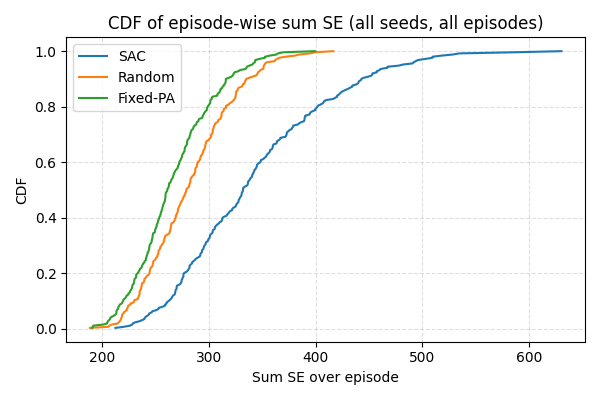}
    \caption{CDF of episode-wise Sum SE. SAC demonstrates superior tail distribution performance.}
    \label{fig:cdf}
\end{figure}

In this section, we evaluate the performance of the proposed SAC-based dynamic beamforming algorithm. The simulation scenario considers a $100 \times 100$ m$^2$ UMi service area with $N=4$ parallel waveguides deployed at a height of $H=10$ m. The system operates at a carrier frequency of $28$ GHz. There are $K=3$ mobile users following the Gauss-Markov mobility model with a maximum velocity of $1.2$ m/s. The four waveguides are deployed by the set $\mathcal{Y}_{\text{WG}} = \{20m, 40m,60m,80m\}$. The optimization decision interval is set to $\Delta t = 4$ s, during which each PA is constrained to slide a maximum distance of $\Delta_{\max} = 3$ m. The energy constraint $P_{\mathrm{max}}$ of the system at CC is normalized and its value is fixed to 1. The detailed simulation parameters are listed in Table \ref{tab:hyperparameters}. We compare our proposed method against two baselines:
\begin{itemize}
    \item Random Policy: The PAs move randomly with actions sampled from a uniform distribution $\mathbf{a}_t \sim \mathcal{U}(-1, 1)$, subject to the maximum sliding constraint.
    \item Fixed-PA: The PAs are stationary and fixed at the center of the waveguides ($x=50$ m) throughout the service duration.
\end{itemize}
All statistical results are averaged over 5 different random seeds to ensure reliability.

\subsection{Overall Performance Analysis}

Fig. \ref{fig:seeds} compares the average performance across different initialization seeds. The SAC algorithm consistently outperforms both baselines across all tested scenarios, demonstrating strong robustness against environmental initialization. Fig. \ref{fig:cdf} further presents the cumulative distribution function (CDF) of the episode-wise sum SE. 


Table \ref{tab:results_table} summarizes the quantitative results. The proposed SAC algorithm achieves an average sum SE of 343.49, yielding a significant gain of 28.6\% over the fixed-PA benchmark.

\begin{table}[htbp]
\centering
\caption{Quantitative Performance Summary (Averaged over 5 Seeds)}
\label{tab:results_table}
\begin{tabular}{l c c c}
\toprule
\textbf{Method} & \textbf{Mean Sum SE} & \textbf{Std. Dev.} & \textbf{Improvement} \\
\midrule
Fixed-PA & 267.16 & $\pm 2.66$ & - \\
Random Policy & 282.10 & $\pm 8.68$ & +5.6\% \\
\textbf{Proposed SAC} & \textbf{343.49} & $\pm \mathbf{12.10}$ & \textbf{+28.6\%} \\
\bottomrule
\end{tabular}
\end{table}



\subsection{Case Study: Spatiotemporal Tracking}

\begin{figure*}[t]
    \centering
    \begin{minipage}{0.24\textwidth}
        \centering
        \includegraphics[width=\linewidth]{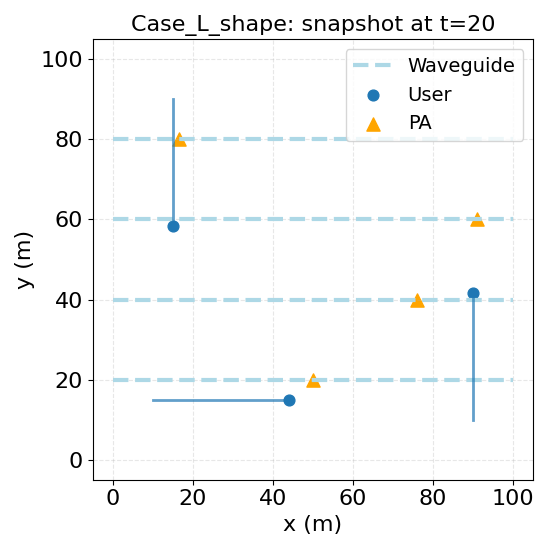}
        \vspace{2pt}
        \\(a) $t=20$ (Pos) 
    \end{minipage}
    \hfill
    \begin{minipage}{0.24\textwidth}
        \centering
        \includegraphics[width=\linewidth]{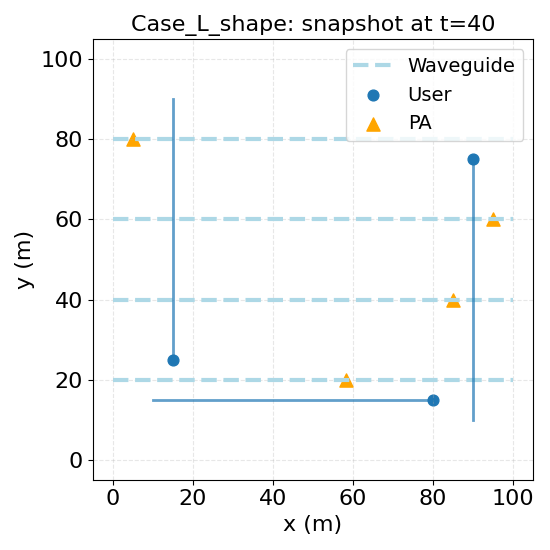}
        \vspace{2pt}
        \\(b) $t=40$ (Pos) 
    \end{minipage}
    \hfill
    \begin{minipage}{0.24\textwidth}
        \centering
        \includegraphics[width=\linewidth]{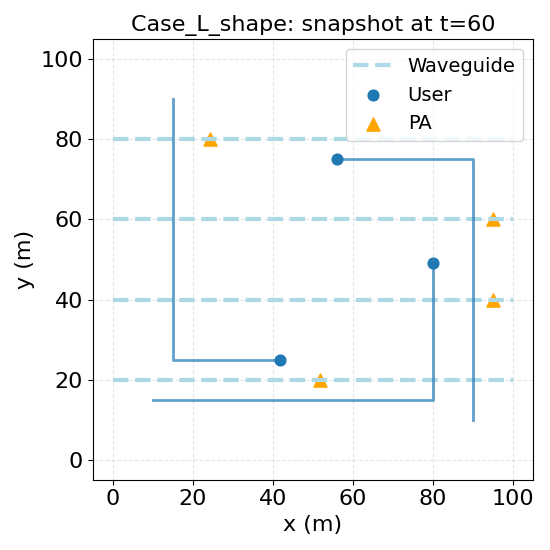}
        \vspace{2pt}
        \\(c) $t=60$ (Pos) 
    \end{minipage}
    \hfill
    \begin{minipage}{0.24\textwidth}
        \centering
        \includegraphics[width=\linewidth]{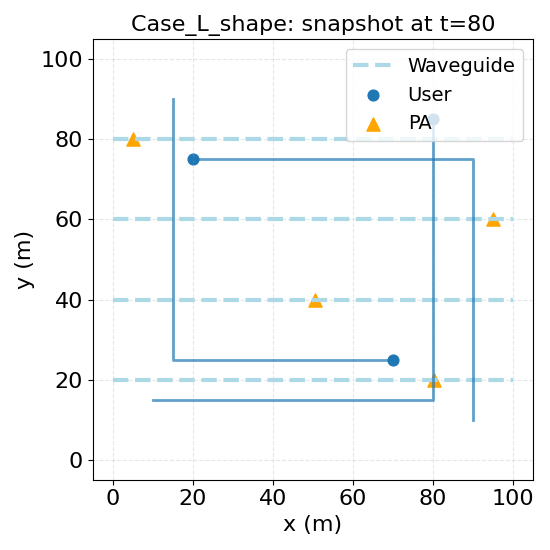}
        \vspace{2pt}
        \\(d) $t=80$ (Pos) 
    \end{minipage}
    
    \vspace{0.3cm} 
    
    \begin{minipage}{0.48\textwidth}
        \centering
        \includegraphics[width=\linewidth]{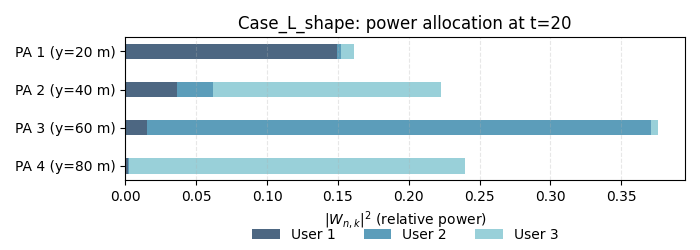}
        \vspace{2pt}
        \\(e) $t=20$ (Power Allocation) 
    \end{minipage}
    \hfill
    \begin{minipage}{0.48\textwidth}
        \centering
        \includegraphics[width=\linewidth]{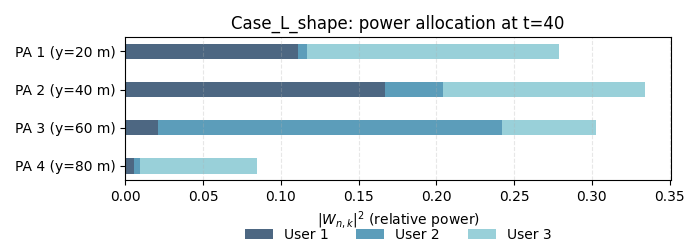}
        \vspace{2pt}
        \\(f) $t=40$ (Power Allocation) 
    \end{minipage}
    
    \vspace{0.2cm} 
    
    \begin{minipage}{0.48\textwidth}
        \centering
        \includegraphics[width=\linewidth]{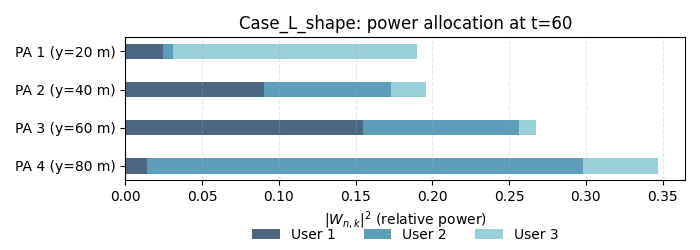}
        \vspace{2pt}
        \\(g) $t=60$ (Power Allocation) 
    \end{minipage}
    \hfill
    \begin{minipage}{0.48\textwidth}
        \centering
        \includegraphics[width=\linewidth]{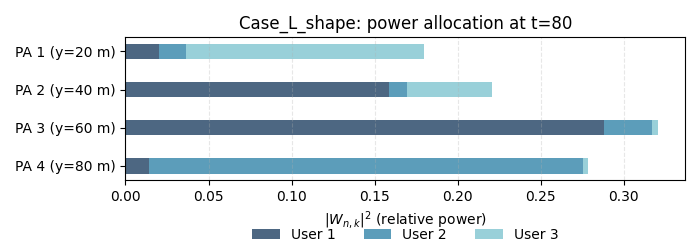}
        \vspace{2pt}
        \\(h) $t=80$ (Power Allocation) 
    \end{minipage}
    
    \caption{Case study of autonomous user tracking. (a)-(d): Dynamic snapshots of PAs (orange) and users (blue) showing proactive geometric adjustment. (e)-(h): Real-time power allocation $\|\mathbf{W}_{n,k}\|^2$. The visualization confirms the system's ability to maintain reliable links through coupled antenna positioning and digital precoding.}
    \label{fig:case_study}
\end{figure*}

To visualize the tracking capability, we simulate a deterministic "L-shape" trajectory that mimics street-level mobility in UMi. Figs. \ref{fig:case_study}(a)-(d) demonstrate that the trained agent has successfully learned the underlying environmental dynamics. The pinching antennas proactively track the users to maintain continuous LoS links, effectively creating a "shadowing" effect while strictly adhering to mechanical energy constraints. Complementing this, the power allocation profiles in Figs. \ref{fig:case_study}(e)-(h) confirm that each user is consistently served by at least one dominant antenna at any given instance. This ensures robust connectivity and validates the system's reliability throughout the dynamic process.

\section{Conclusion}
This paper proposed a bilevel optimization framework for dynamic antenna positioning and beamforming precoding in UMi PASS networks with multiple mobile users. In the proposed framework, an SAC agent is designed to dynamically optimize PA positions, while ZF precoding is adopted for low-complexity beamforming computation. Numerical results demonstrate that the agent autonomously learns strategic positioning to mitigate blockages, achieving a 28.6\% spectral efficiency gain over fixed-antenna systems. These findings validate the potential of DRL-driven flexible hardware for robust UMi communications, laying the groundwork for future research on energy-aware and multi-cell cooperative networks.


%

\appendices

\ifCLASSOPTIONcaptionsoff
  \newpage
\fi



%

\bibliographystyle{IEEEtran}  
\bibliography{reference_new}   

\begin{thebibliography}{10}
\providecommand{\url}[1]{#1}
\csname url@samestyle\endcsname
\providecommand{\newblock}{\relax}
\providecommand{\bibinfo}[2]{#2}
\providecommand{\BIBentrySTDinterwordspacing}{\spaceskip=0pt\relax}
\providecommand{\BIBentryALTinterwordstretchfactor}{4}
\providecommand{\BIBentryALTinterwordspacing}{\spaceskip=\fontdimen2\font plus
\BIBentryALTinterwordstretchfactor\fontdimen3\font minus
  \fontdimen4\font\relax}
\providecommand{\BIBforeignlanguage}[2]{{%
\expandafter\ifx\csname l@#1\endcsname\relax
\typeout{** WARNING: IEEEtran.bst: No hyphenation pattern has been}%
\typeout{** loaded for the language `#1'. Using the pattern for}%
\typeout{** the default language instead.}%
\else
\language=\csname l@#1\endcsname
\fi
#2}}
\providecommand{\BIBdecl}{\relax}
\BIBdecl

\bibitem{guo2025gpass}
J.~Guo, Y.~Liu, and A.~Nallanathan, ``{GPASS}: Deep learning for beamforming in
  pinching-antenna systems ({PASS}),'' \emph{arXiv preprint arXiv:2502.01438},
  2025.

\bibitem{ding2025flexible}
Z.~Ding, R.~Schober, and H.~V. Poor, ``Flexible-antenna systems: A
  pinching-antenna perspective,'' \emph{IEEE Transactions on Communications},
  2025.

\bibitem{ouyang2025array}
C.~Ouyang, Z.~Wang, Y.~Liu, and Z.~Ding, ``Array gain for pinching-antenna
  systems ({PASS}),'' \emph{IEEE Communications Letters}, 2025.

\bibitem{wang2025antenna}
K.~Wang, Z.~Ding, and R.~Schober, ``Antenna activation for {NOMA} assisted
  pinching-antenna systems,'' \emph{IEEE Wireless Communications Letters},
  2025.

\bibitem{zhao2025pinching}
J.~Zhao, H.~Song, X.~Mu, K.~Cai, Y.~Zhu, and Y.~Liu, ``Pinching-antenna
  systems-enabled multi-user communications: Transmission structures and
  beamforming optimization,'' \emph{arXiv preprint arXiv:2508.14458}, 2025.

\bibitem{zhou2025gradient}
K.~Zhou, W.~Zhou, D.~Cai, X.~Lei, Y.~Xu, Z.~Ding, and P.~Fan, ``A gradient
  meta-learning joint optimization for beamforming and antenna position in
  pinching-antenna systems,'' \emph{arXiv preprint arXiv:2506.12583}, 2025.

\bibitem{zeng2025robust}
M.~Zeng, X.~Wang, Y.~Liu, Z.~Ding, G.~K. Karagiannidis, and H.~V. Poor,
  ``Robust resource allocation for pinching-antenna systems under imperfect
  {CSI},'' \emph{arXiv preprint arXiv:2507.12582}, 2025.

\bibitem{xie2025low}
X.~Xie, F.~Fang, Z.~Ding, and X.~Wang, ``A low-complexity placement design of
  pinching-antenna systems,'' \emph{IEEE Communications Letters}, 2025.

\bibitem{qin2025physics}
H.~Qin, Z.~Wu, Y.~Liu, X.~Zhang, and X.~Zhang, ``Physics-based trajectory
  design for cellular-connected {UAV} in rainy environments based on deep
  reinforcement learning,'' \emph{IEEE Transactions on Intelligent
  Transportation Systems}, 2025.

\bibitem{zhang2021cognitive}
B.~Zhang, C.~Jin, K.~Cao, Q.~Lv, and R.~Mittra, ``Cognitive conformal antenna
  array exploiting deep reinforcement learning method,'' \emph{IEEE
  Transactions on Antennas and Propagation}, vol.~70, no.~7, pp. 5094--5104,
  2021.

\bibitem{wu2025intelligent}
K.~Wu, Q.~Zhao, Z.~Feng, Y.~Mu, H.~Qin, X.~Zhang, and X.~Zhang, ``Intelligent
  optimization of wireless access point deployment for communication-based
  train control systems using deep reinforcement learning,'' \emph{arXiv
  preprint arXiv:2509.24819}, 2025.

\bibitem{zhu20213gpp}
Q.~Zhu, C.-X. Wang, B.~Hua, K.~Mao, S.~Jiang, and M.~Yao, ``{3GPP} {TR} 38.901
  channel model,'' in \emph{the {wiley} 5G Ref: the essential 5G reference
  online}.\hskip 1em plus 0.5em minus 0.4em\relax Wiley Press, 2021, pp. 1--35.

\end{thebibliography}

%





\end{document}